\documentclass[runningheads]{llncs}
\usepackage{graphicx}
\usepackage[table]{xcolor}
\usepackage{multirow}
\usepackage{wrapfig}

\begin{document}
\title{Your Tribe Decides Your Vibe: Analyzing Local Popularity in the US Patent Citation Network}
%
%
\author{Nishit Narang\inst{1} \and
Manoj Kumar Ganji\inst{1} \and
Amit Anil Nanavati\inst{1,2}}
\authorrunning{N. Narang et al.}
%
\institute{Indian Institue of Technology, Delhi, New Delhi, India \and
IBM GTS Innovation Labs, New Delhi. India
\email{\{nisnarang,manojganji645\}@gmail.com}\\
\email{nanavati@acm.org}}
\maketitle              
\begin{abstract}
In many networks, the indegree of a vertex is a measure of its popularity. Past research has studied indegree distributions treating the network as a whole. In the US Patent citation network (USPCN), patents are classified into categories and subcategories. A natural question arises: How do patents gather their popularity from various (sub)categories? We analyse local indegree distributions to answer this question.

The citation (indegree) of a patent within the same category indicates its internal popularity, while a cross-category citation indicates its external popularity. We analyze the internal and external indegree distributions at each level of USPCN hierarchy to learn how the internal and external popularity of patents varies across (sub)categories.

We find that all (sub)categories have local preferences that decide internal and external patents' popularities. Different patents are popular in different groups: Groups C1, C2 and C3 may not agree on popular patents in C1. In general, patent popularity appears to be a highly local phenomenon with subcategories (not even categories) deciding their own popular patents independent of the other (sub)categories.

\end{abstract}
\section{Introduction}

Social network formation is a complex process in which entities (nodes) simultaneously attempt to satisfy their goals under multiple, possibly conflicting, constraints [1]. The challenge in social network analysis is in the identification of local processes of link formation that lead to the global properties of the network.

In the study of real-world networks, degree distribution is one of the fundamental properties. Several networks exhibit a power-law in their degree distributions. The preferential attachment model introduced in [2] proposed a generative model capable of explaining evolution of networks with power law degree distributions.

While there have been many studies to understand global network evolution, there have been few studies which have attempted to model local behaviors in networks with predefined categories or groups. As such, there is limited understanding of how local subgraph processes generate or produce global graph properties. A model for networks with communities is the Stochastic Block Model (SBM) [3], which generates networks with uniform or homogeneous degree distributions. To mimic power law distributions of real-world networks, SBM variants [4,5] have been proposed. However, none of the SBM variants account for power law degree distributions at local (group or community) levels. Recent research attempts to study the effect of local events and information hiding on the overall network evolution. A few studies on this topic are discussed in [8], [9], [10] and [11]. Most of these studies describe the occurrence of local events in randomly selected subgraphs, but not in the context of groups or communities. We discuss more about these in the related work in Section 4.

In this paper, our goal is to investigate networks with explicit, predefined (ground-truth) categories or groups, with an aim to understand how local indegree distributions aggregate to generate the global indegree distribution. By local, we mean the degree distribution of vertices within a group (internal degree, within group edges) as well as across groups (external degree, based on cross-group links). We selected the USPCN dataset for its predefined categories and subcategories. Since the categories are prede-fined (unlike communities based on modularity for example), this gives us a ground-truth for our investigation. We study these local distributions across these two levels of hierarchy. The USPCN is a single-membership network, where each vertex in the network belongs to exactly one (sub)category. The novelty of our research is the analysis of local degree distributions at subgraph hierarchies, in order to understand the local processes behind link formation. In our study of local distributions within USPCN, we attempt to find an answer to the following questions:
\begin{itemize}
\item Is patent popularity a result of global processes or local processes?
\item How does internal and external popularity vary?
\item Are internally popular vertices also externally popular?
\item What fraction of vertices are externally popular?
\item How does the popularity of a vertex vary from one external group to another?
\item How does popularity vary across the different levels of hierarchy in the network? 
\item How do local indegree distributions aggregate to the global indegree distribution?
\end{itemize}

The findings (Section 3.2) to above questions help us define a local behavior-driven generative model for the US patents citation network (Section 3.3).

\section{Data Description}

In our research, we study the US Patents Citation Network [12], with its predefined categories. The US Patents dataset includes information on patents granted over a period of 37 years until Dec 30, 1999. The citation graph is however available for only the latter 25 years, starting from year 1975, totaling 16,522,438 citations. In our study, we exclude patents that were not defined to fall into any of the predefined categories, thereby resulting in a reduction of the total citations used in our study to 13,968,440 citations (i.e. an approximate 85\% of the full network edges). A description of the tiered group structure for the US patents dataset is provided in Table 1. Fig. 1 shows the USPCN global network properties, with the indegree distribution following a power law with a gamma value of 4.46. The figure also provides a visual representation of the top-tier structure of the USPCN.

\setlength{\tabcolsep}{2pt}
\begin{table}
\centering
    \caption{US Patents Dataset: Tiered Community Structure. The brackets contain the \#nodes.} 
    \label{tab:}
\begin{tabular}{|l|l|l|} 
\hline
\rowcolor{lightgray} {\bf 1. Chemical (457353)}  & {\bf 2. Computers \&}  & {\bf 3. Drugs \& Medical}             \\
\rowcolor{lightgray} & {\bf Communications (218942)} & {\bf (139524)}\\ \hline
11. Agriculture, Food,                  & 21. Communications               & 31. Drugs (52586) \\
    Textiles (18571)  & (96119)& \\ \hline
12. Coating (34443)                     & 22. Computer HW \& SW            & 32. Surgery \& Med Inst.  \\ 
& (67186)& (56503)\\ \hline
13. Gas (12101)                         & 23. Computer Peripherals        & 33. Biotechnology     \\ 
&   (17419) & (17392)  \\ \hline
14. Organic Compounds           & 24.   Information Storage        & 39.   Miscellaneous --  \\
 (79947) & (38218)&  Drugs \& Med (13043) \\ \hline
15.   Resins (77051)                    & \cellcolor{lightgray} {\bf 5. Mechanical (530675)} & \cellcolor{lightgray} {\bf 6. Others (505859)}                       \\ \hline
19.   Miscellaneous-chemical   & 51. Mat. Proc \& Handling       & 61. Agriculture,  \\
 (235240) & (131787) & Husbandry, Food (50350)\\ \hline
\cellcolor{lightgray}{\bf 4.Electrical \& Electronic}      & 52. Metal Working (70511)               & 62. Amusement Devices             \\
\cellcolor{lightgray} {\bf (387996)} & (23237) & \\ \hline
41. Electrical Devices          & 53. Motors \& Engines +  & 63. Apparel \& Textile           \\
 (75841) & Parts (87574) & (41523) \\ \hline 
42. Electrical Lighting          & 54. Optics (50088)                      & 64. Earth Working \&         \\
(36243) & & Wells (35976) \\ \hline 
43. Measuring \& Testing         & 55. Transportation (69593)              & 65. Furniture, House      \\
(66093) & & Fixtures (48658) \\ \hline 
44. Nuclear \& X-rays    & 59.   Miscellaneous --   & 66. Heating (31959)      \\
(34132) & Mechanical (121122) & \\ \hline 
45. Power Systems (81754)       &              & 67.   Pipes \& Joints \\
& & (22216) \\ \cline{1-1}\cline{3-3}
46. Semiconductor Devices        &                      & 68.   Receptacles (51905)  \\
(39439) & &  \\ \cline{1-1}\cline{3-3}
49.   Miscellaneous-Elec        &                       & 69.   Miscellaneous-Others  \\    
(54494) & & (200035) \\ \hline 
\end{tabular}
\end{table}

\begin{figure}
\includegraphics[width=\textwidth]{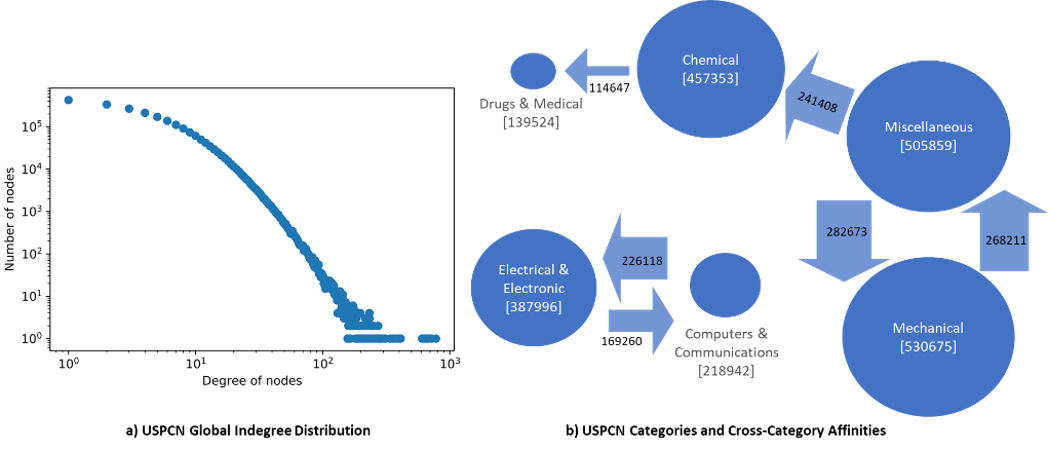}
\caption{USPCN Global Properties. Notice a) the power law in the global indegree distribution and b) the disparity in sizes and affinity relationship of categories.} \label{fig1}
\end{figure}

The US patents network has six predefined top-level categories, which we treat as the top-tier categorization of patents. We define the Tier-1 hierarchy as consisting of the overall network and its decomposition into categories. Each category is further sub-divided into subcategories. The subcategories are the Tier-2 hierarchy of the citation network. The Edge matrix for the USPCN (or the SBM matrix) is depicted in Fig. 2 and Fig. 3, for categories and subcategories respectively.

\begin{figure}
\includegraphics[width=\textwidth]{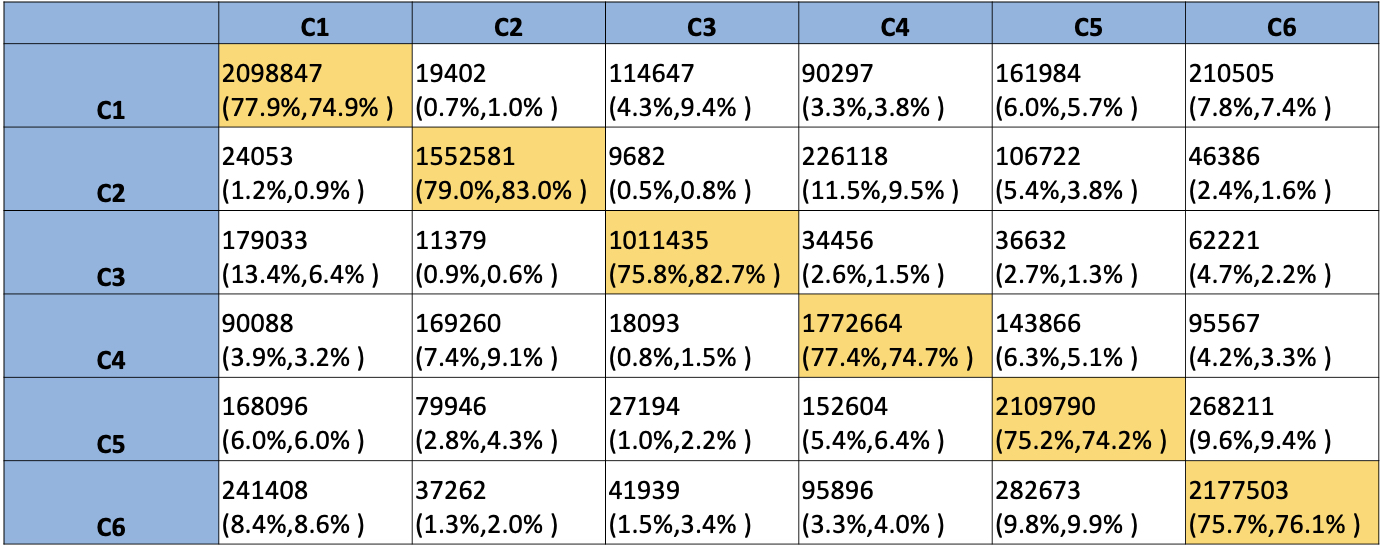}
\caption{Citation Edge Matrix for US Patent Dataset. Notice that the diagonal elements are an order of magnitude higher than the off-diagonal values.} \label{fig2}
\end{figure}

\begin{figure}
\includegraphics[width=\textwidth]{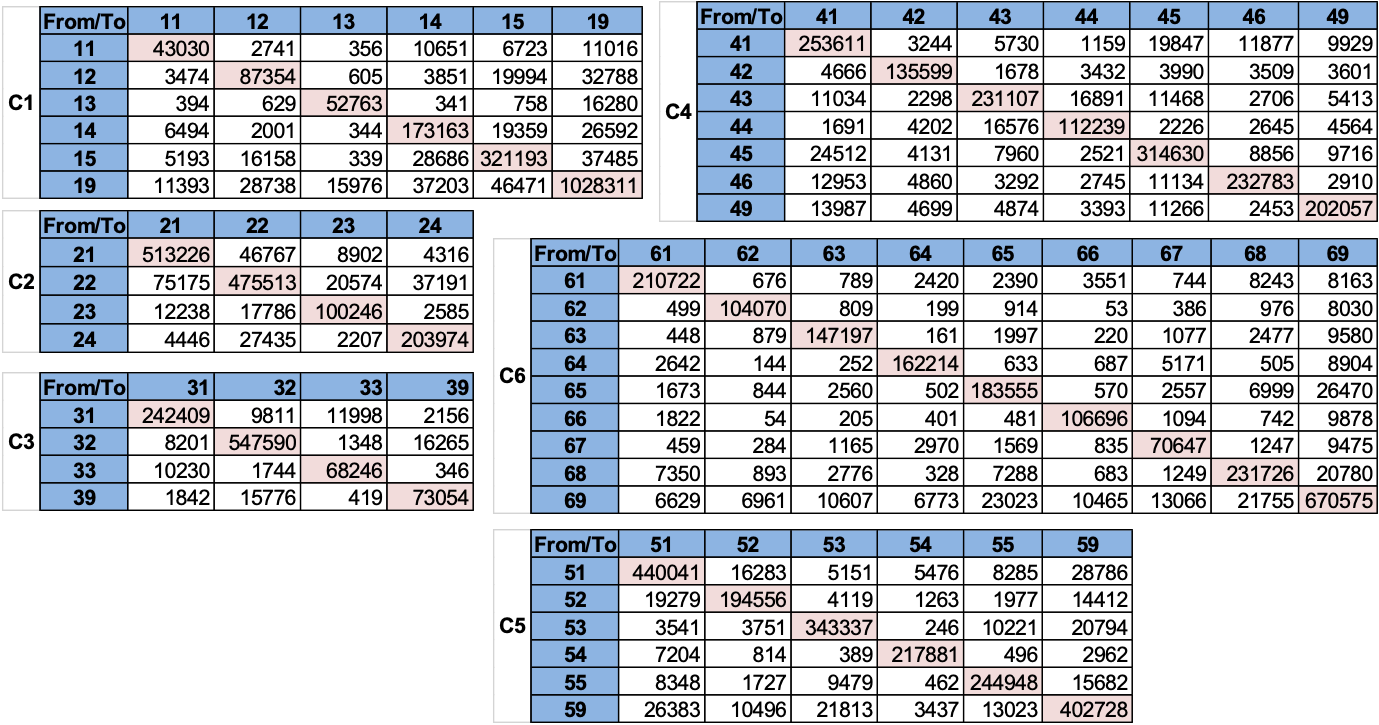}
\caption{Citation Edge Matrix for Categories within US Patent Dataset. Notice the disparity between diagonal and off-diagonal values.} \label{fig3}
\end{figure}

\section{Analysis of Local Distributions}

While most studies and models focus on the {\bf global} degree distribution of networks, they overlook the variety and patterns present at local levels, which we study for the first time in this paper. We believe that for category-based networks, it is possible and important to study the local distributions, in order to understand how the local indegree distributions aggregate to the global. The two-tiered hierarchical structure of the USPCN allows us to perform analysis of local indegree distributions at different levels within the network. At each level as well as across levels, we perform the following tests:
\begin{itemize}
  \item Separation of internal and external degree distributions from the global degree distribution. This is done to extract the local distributions from the aggregate.
  \item Pair-wise calculation of Pearson correlation coefficient (PCC) between global, internal and external degree of vertices. This is done to understand variation in vertex popularity internally and externally.
  \item Likelihood Ratio Test (LRT) for global, internal and external degree distributions, using the powerlaw package [13], to find the best-fit distribution type from amongst power law (PL), truncated power law (TPL), exponential (EXP), stretched exponential (SE), lognormal (LN) and lognormal positive (LNP).
    \item Using Pearson correlation coefficient (PCC) to measure the similarity or variation in the popularity of vertices from one group to another.
\end{itemize}

\subsection{Experiments and Observations}

Likelihood Ratio Test (LRT) on the indegree distribution in the global USPCN reveals Truncated Power Law as the best-fit distribution (Fig. 1a). Results at sub-hierarchical levels demonstrate subtle variations and are captured in Table 2 and Table 4 for categories and subcategories respectively. Pair-wise degree correlations (using PCC) between the global, internal and external degrees are captured in Table 3 and Table 4 for categories and subcategories respectively. Also, for each (sub-)category, the percentage of vertices that have in-links from external (sub-)categories is depicted in Table 2 (for Tier-1) and Table 4 (for Tier-2).


\begin{table}
\centering
    \caption{Tier-1 (Category-level) Degree Distribution LRT Results}
\begin{tabular}{ccccc}
Group /& Global & Internal & \% of externally & External Distribution \\ 
Category & Distribution & Distribution & popular vertices & (from each external group) \\\hline
1                & TPL                 & TPL                   & 28\%                              & TPL, PL                                          \\
2                & TPL                 & TPL                   & 26\%                              & TPL, PL*                                         \\
3                & TPL                 & TPL                   & 27\%                              & TPL, PL                                          \\
4                & TPL                 & TPL                   & 28\%                              & TPL, PL                                          \\
5                & TPL                 & TPL                   & 25\%                              & TPL, PL                                          \\
6                & TPL                 & TPL                   & 25\%                              & TPL, PL \\ \hline                               
\end{tabular}
\end{table}
\vspace{-0.3in}
{\footnotesize
\noindent TPL -- Truncated Power Law; PL -- Power Law \\
$^{*}$An Exception is the external degree distribution from category 6 to category 2, where lognormal distribution was also a joint best-fit along-with TPL and PL.\\
}

Tier-2 analysis was executed on each of the six top-level categories, treating each group of nodes and edges as an independent graph. For each subcategory in a category, the external in-links only included links coming from other subcategories from {\bf within} the same category.

\begin{table}
\centering
    \caption{Tier-1 (Category-level) Degree Correlation Results}
\footnotesize
\begin{tabular}{cccc}
\hline
Group/ & \multicolumn{3}{c}{Degree Correlation (PCC)} \\  \cline{2-4}
Category        & Global – Internal & Global – External & Internal – External \\ \hline
1                                                                                & 0.91              & 0.64              & 0.25                \\
2                                                                                & 0.96              & 0.45              & 0.18                \\
3                                                                                & 0.97              & 0.52              & 0.28                \\
4                                                                                & 0.9               & 0.65              & 0.25                \\
5                                                                                & 0.86              & 0.66              & 0.19                \\
6                                                                                & 0.88              & 0.62              & 0.18              \\ \hline  
\end{tabular}
\end{table}


\begin{table}
\centering
    \caption{Tier-2 (Subcategory) Degree Distribution LRT and Degree Correlation Results}
\footnotesize
\begin{tabular}{cccccccc}
\hline
Subcat- & Global  & Internal & External Distri-        & \% of exter-  & \multicolumn{3}{c}{Degree Correlation (PCC)}  \\ \cline{6-8}
egory   & Distri- & Distri-  & bution (from all ext-   & nally popular &  Global -- & Global --  & Internal -- \\
        & bution  & bution   & ernal subcategories)    & vertices      &  Internal  & External   & External  \\ \hline
11                           & TPL                                  & TPL                                    & TPL, LN                                                                  & 27\%                                               & 0.8               & 0.75              & 0.2                 \\
12                           & TPL                                  & TPL                                    & TPL                                                                      & 28\%                                               & 0.82              & 0.74              & 0.22                \\
13                           & TPL                                  & TPL                                    & TPL                                                                      & 29\%                                               & 0.92              & 0.57              & 0.21                \\
14                           & TPL                                  & TPL                                    & TPL, PL, LN                                                              & 18\%                                               & 0.82              & 0.72              & 0.19                \\
15                           & TPL                                  & TPL                                    & TPL, PL                                                                  & 24\%                                               & 0.94              & 0.64              & 0.33                \\
19                           & TPL                                  & TPL                                    & TPL, PL                                                                  & 10\%                                               & 0.96              & 0.46              & 0.2                 \\
21                           & TPL                                  & TPL                                    & TPL                                                                      & 16\%                                               & 0.95              & 0.54              & 0.26                \\
22                           & TPL                                  & TPL                                    & TPL                                                                      & 23\%                                               & 0.96              & 0.6               & 0.36                \\
23                           & TPL                                  & TPL                                    & TPL                                                                      & 27\%                                               & 0.96              & 0.59              & 0.34                \\
24                           & TPL                                  & TPL                                    & TPL                                                                      & 17\%                                               & 0.91              & 0.61              & 0.22                \\
31                           & TPL                                  & TPL                                    & TPL, PL, LN                                                              & 7\%                                                & 0.98              & 0.49              & 0.31                \\
32                           & SE                                   & SE                                     & TPL, PL                                                                  & 7\%                                                & 0.98              & 0.37              & 0.17                \\
33                           & TPL                                  & TPL                                    & TPL, PL                                                                  & 14\%                                               & 0.96              & 0.51              & 0.24                \\
39                           & TPL                                  & TPL                                    & TPL                                                                      & 19\%                                               & 0.9               & 0.79              & 0.43                \\
41                           & TPL                                  & TPL                                    & TPL, PL, LN                                                              & 18\%                                               & 0.91              & 0.51              & 0.09                \\
42                           & TPL                                  & TPL                                    & TPL, PL                                                                  & 13\%                                               & 0.96              & 0.46              & 0.2                 \\
43                           & TPL                                  & TPL                                    & TPL, PL                                                                  & 12\%                                               & 0.95              & 0.55              & 0.24                \\
44                           & TPL                                  & TPL                                    & TPL, PL                                                                  & 18\%                                               & 0.91              & 0.5               & 0.1                 \\
45                           & TPL                                  & TPL                                    & TPL, PL                                                                  & 14\%                                               & 0.92              & 0.51              & 0.15                \\
46                           & TPL                                  & TPL                                    & TPL, PL, LN                                                              & 16\%                                               & 0.97              & 0.41              & 0.16                \\
49                           & TPL                                  & TPL                                    & TPL, PL, LN                                                              & 13\%                                               & 0.95              & 0.33              & 0.01                \\
51                           & TPL                                  & TPL                                    & TPL, PL                                                                  & 10\%                                               & 0.95              & 0.43              & 0.12                \\
52                           & TPL                                  & TPL                                    & TPL, PL                                                                  & 10\%                                               & 0.94              & 0.42              & 0.08                \\
53                           & TPL                                  & TPL                                    & TPL, PL                                                                  & 10\%                                               & 0.96              & 0.33              & 0.06                \\
54                           & TPL                                  & TPL                                    & TPL, PL                                                                  & 4\%                                                & 0.98              & 0.32              & 0.12                \\
55                           & TPL                                  & TPL                                    & TPL, PL                                                                  & 10\%                                               & 0.95              & 0.39              & 0.08                \\
59                           & TPL                                  & TPL                                    & TPL, PL                                                                  & 14\%                                               & 0.95              & 0.38              & 0.07                \\
61                           & TPL                                  & TPL                                    & TPL, PL                                                                  & 7\%                                                & 0.91              & 0.52              & 0.11                \\
62                           & TPL                                  & TPL                                    & TPL, PL                                                                  & 9\%                                                & 0.97              & 0.29              & 0.04                \\
63                           & TPL                                  & TPL                                    & LN                                                                       & 9\%                                                & 0.95              & 0.39              & 0.08                \\
64                           & TPL                                  & TPL                                    & TPL, PL                                                                  & 8\%                                                & 0.98              & 0.29              & 0.07                \\
65                           & TPL                                  & TPL                                    & LN                                                                       & 16\%                                               & 0.92              & 0.44              & 0.07                \\
66                           & TPL                                  & TPL                                    & TPL, PL, LN                                                              & 11\%                                               & 0.94              & 0.37              & 0.05                \\
67                           & TPL                                  & TPL                                    & TPL, PL, LN                                                              & 23\%                                               & 0.88              & 0.51              & 0.04                \\
68                           & TPL                                  & TPL                                    & TPL, PL, LN                                                              & 16\%                                               & 0.95              & 0.58              & 0.29                \\
69                           & TPL                                  & TPL                                    & TPL, PL                                                                  & 10\%                                               & 0.94              & 0.46              & 0.13
\\ \hline
\end{tabular}
\hspace{-0.2in}{\footnotesize PL -- Power Law; TPL -- Truncated PL; LN -- LogNormal; SE -- Stretched Exponential}
\end{table}

Table 5 depicts the pair-wise external degree correlations from two external source categories into the same destination category. This is indicative of how the popularity of vertices within the same (destination) category varies based on the external (source) category.

\begin{table}
\vspace{-0.2in}
\centering
    \caption{Tier-1 (Category-level) Pairwise External Degree Correlation Results}
\scriptsize
\begin{tabular}{cccc||ccc}
\hline
From Cat.     & From Cat.    & Degree Corr.              &  & From  Cat.    & From Cat.    & Degree Corr.     \\          
(a)       & (b)      & (PCC) [(a) and (b)]  &  & (a)       & (b)      & (PCC) [(a) and (b)] \\ \hline
\multicolumn{3}{c}{To Category: 1}                                                                                                                                                                                                                &  & \multicolumn{3}{c}{To Category: 4}                                                                                                                                                                                                                      \\ \hline
2                                                                     & 3                                                                        & -0.13                                                                                          &  & 1                                                                        & 2                                                                     & -0.13                                                                                                \\
2                                                                     & 4                                                                        & -0.08                                                                                          &  & 1                                                                        & 3                                                                     & -0.19                                                                                                \\
2                                                                     & 5                                                                        & -0.05                                                                                          &  & 1                                                                        & 5                                                                     & -0.12                                                                                                \\
2                                                                     & 6                                                                        & -0.19                                                                                          &  & 1                                                                        & 6                                                                     & -0.22                                                                                                \\
3                                                                     & 4                                                                        & -0.18                                                                                          &  & 2                                                                        & 3                                                                     & -0.07                                                                                                \\
3                                                                     & 5                                                                        & -0.18                                                                                          &  & 2                                                                        & 5                                                                     & -0.13                                                                                                \\
3                                                                     & 6                                                                        & -0.15                                                                                          &  & 2                                                                        & 6                                                                     & -0.13                                                                                                \\
4                                                                     & 5                                                                        & -0.11                                                                                          &  & 3                                                                        & 5                                                                     & -0.14                                                                                                \\
4                                                                     & 6                                                                        & -0.21                                                                                          &  & 3                                                                        & 6                                                                     & -0.19                                                                                                \\
5                                                                     & 6                                                                        & -0.14                                                                                          &  & 5                                                                        & 6                                                                     & -0.18                                                                                                \\ \hline
\multicolumn{3}{c}{To Category: 2}                                                                                                                                                                                                                &  & \multicolumn{3}{c}{To Category: 5}                                                                                                                                                                                                                      \\ \hline
1                                                                     & 3                                                                        & -0.14                                                                                          &  & 1                                                                        & 2                                                                     & -0.19                                                                                                \\
1                                                                     & 4                                                                        & -0.05                                                                                          &  & 1                                                                        & 3                                                                     & -0.07                                                                                                \\
1                                                                     & 5                                                                        & -0.04                                                                                          &  & 1                                                                        & 4                                                                     & -0.17                                                                                                \\
1                                                                     & 6                                                                        & -0.17                                                                                          &  & 1                                                                        & 6                                                                     & -0.16                                                                                                \\
3                                                                     & 4                                                                        & -0.02                                                                                          &  & 2                                                                        & 3                                                                     & -0.12                                                                                                \\
3                                                                     & 5                                                                        & -0.13                                                                                          &  & 2                                                                        & 4                                                                     & -0.07                                                                                                \\
3                                                                     & 6                                                                        & -0.21                                                                                          &  & 2                                                                        & 6                                                                     & -0.20                                                                                                \\
4                                                                     & 5                                                                        & -0.09                                                                                          &  & 3                                                                        & 4                                                                     & -0.12                                                                                                \\
4                                                                     & 6                                                                        & -0.09                                                                                          &  & 3                                                                        & 6                                                                     & -0.07                                                                                                \\
5                                                                     & 6                                                                        & -0.17                                                                                          &  & 4                                                                        & 6                                                                     & -0.23                                                                                                \\ \hline
\multicolumn{3}{c}{To Category: 3}                                                                                                                                                                                                                &  & \multicolumn{3}{c}{To Category: 6}                                                                                                                                                                                                                      \\ \hline
1                                                                     & 2                                                                        & -0.14                                                                                          &  & 1                                                                        & 2                                                                     & -0.20                                                                                                \\
1                                                                     & 4                                                                        & -0.18                                                                                          &  & 1                                                                        & 3                                                                     & -0.05                                                                                                \\
1                                                                     & 5                                                                        & -0.10                                                                                          &  & 1                                                                        & 4                                                                     & -0.19                                                                                                \\
1                                                                     & 6                                                                        & -0.13                                                                                          &  & 1                                                                        & 5                                                                     & -0.09                                                                                                \\
2                                                                     & 4                                                                        & -0.04                                                                                          &  & 2                                                                        & 3                                                                     & -0.19                                                                                                \\
2                                                                     & 5                                                                        & -0.16                                                                                          &  & 2                                                                        & 4                                                                     & -0.12                                                                                                \\
2                                                                     & 6                                                                        & -0.21                                                                                          &  & 2                                                                        & 5                                                                     & -0.14                                                                                                \\
4                                                                     & 5                                                                        & -0.21                                                                                          &  & 3                                                                        & 4                                                                     & -0.20                                                                                                \\
4                                                                     & 6                                                                        & -0.23                                                                                          &  & 3                                                                        & 5                                                                     & -0.11                                                                                                \\
5                                                                     & 6                                                                        & -0.15                                                                                          &  & 4                                                                        & 5                                                                     & -0.15 \\ \hline                                                                                              
\end{tabular}
\vspace{-0.3in}
\end{table}
Table 6 depicts LRT results for a random sample study to understand how patents in categories receive citations from external subcategories. This sample study was done to understand the across tier indegree distribution types.


\begin{table}
\centering
    \caption{Sample Degree Distribution LRT Results from Subcategory to Category (across Tiers)}
\footnotesize
\begin{tabular}{ccc}
\hline
To Category & From Subcategory & Distribution Type \\ \hline
4           & 21               & LN                \\
4           & 22               & TPL               \\
6           & 51               & LN                \\
6           & 52               & LN                \\
6           & 53               & TPL, PL, LN       \\
5           & 61               & TPL, PL, LN       \\
5           & 62               & TPL, PL, LN       \\
5           & 69               & TPL, PL, LN       \\
4           & 69               & TPL, PL, LN      \\ \hline 
\end{tabular}
\vspace{0.1in}\\
{\footnotesize PL--Power Law; TPL--Truncated PL; LN--LogNormal}
\end{table}

Table 7 and Table 8 depict the pair-wise external degree correlations from two external source subcategories into the same destination (sub)category. This shows how the popularity of vertices within the same (destination) category/subcategory varies with the external (source) subcategory.

\begin{table}[h]
\centering
    \caption{Sample Degree Correlation Results from Subcategory to Category (across Tiers)}
\begin{tabular}{cccc}
\hline
To Category & From Subcategory & From Subcategory  & Degree Correlation (PCC) \\
            & (a)              & (b)               & (between a and b) \\ \hline
4           & 21                                                                   & 22                                                                   & 0.03                                           \\
4           & 21                                                                   & 69                                                                   & -0.22                                          \\
4           & 22                                                                   & 69                                                                   & -0.21                                          \\
5           & 61                                                                   & 62                                                                   & -0.25                                          \\
5           & 62                                                                   & 69                                                                   & -0.1                                           \\
5           & 61                                                                   & 69                                                                   & -0.15                                          \\
6           & 51                                                                   & 52                                                                   & -0.2                                           \\
6           & 52                                                                   & 53                                                                   & -0.28                                          \\
6           & 51                                                                   & 53                                                                   & -0.28                                   \\ \hline      
\end{tabular}
\end{table}


\begin{table}
\centering
    \caption{Sample Degree Correlation Results from Subcategory to Subcategory}
\begin{tabular}{cccc}
\hline
To Subcategory & From Subcategory & From Subcategory  & Degree Correlation (PCC) \\
            & (a)              & (b)               & (between a and b) \\ \hline
11             & 12                                                                   & 13                                                                   & -0.11                                          \\
11             & 13                                                                   & 14                                                                   & -0.13                                          \\
11             & 14                                                                   & 15                                                                   & -0.2                                           \\
22             & 21                                                                   & 23                                                                   & -0.02                                          \\
22             & 23                                                                   & 24                                                                   & -0.15                                          \\
22             & 21                                                                   & 24                                                                   & -0.13                                          \\
31             & 32                                                                   & 33                                                                   & -0.12                                          \\
31             & 33                                                                   & 39                                                                   & 0.03                                           \\
31             & 32                                                                   & 39                                                                   & 0.06                                           \\
43             & 41                                                                   & 42                                                                   & -0.37                                          \\
43             & 42                                                                   & 44                                                                   & -0.17                                          \\
43             & 41                                                                   & 44                                                                   & -0.31                                          \\
52             & 51                                                                   & 53                                                                   & -0.28                                          \\
52             & 53                                                                   & 54                                                                   & -0.29                                          \\
52             & 51                                                                   & 54                                                                   & -0.05                                          \\
67             & 62                                                                   & 63                                                                   & -0.21                                          \\
67             & 63                                                                   & 64                                                                   & -0.26                                          \\
67             & 64                                                                   & 65                                                                   & -0.31                    \\ \hline                     
\end{tabular}
\end{table}

Based on analysis at both Tier-1 and Tier-2 hierarchies of the graph as well as cross- tier studies, Table 9 summarizes our findings.

\begin{table}
    \centering
    \caption{Analysis of Experimental Results and Related Findings}
\setlength{\tabcolsep}{6pt}
\begin{tabular}{p{8pt}p{4cm}p{7.5cm}}
\hline
\# & \multicolumn{1}{c}{Finding} & \multicolumn{1}{c}{Analysis} \\ \hline
1. &	Local processes are driving local distributions	& Degree correlation between the external and internal indegrees for patents within a (sub)category is low (Tier-1 range: 0.18 to 0.28 [Table 3]; Tier-2 range: 0.01 to 0.43 [Table 4]), indicating that vertices that are internally popular are not the ones which are externally popular. \\ \hline
2. &	Internal popularity of patents has a strong influence on their global popularity &	Degree correlation between the global and internal indegrees for patents within a (sub)category is high (Tier-1 range: 0.86 to 0.97 [Table 3]; Tier-2 range: 0.8 to 0.98 [Table 4]). \\ \hline
3. &	External popularity of patents does contribute to the overall global popularity, albeit to a lesser extent compared to the internal popularity &	Degree correlation between the global and external indegrees for patents within a (sub)category is medium (Tier-1 range: 0.45 to 0.66 [Table 3]; Tier-2 range: 0.29 to 0.79 [Table 4]) \\ \hline
4. &	Patent popularity varies considerably across external groups, and across hierarchical levels, strengthening finding \#1.	& At Tier-1 hierarchy, considering patent citations originating from any two external source categories towards the same destination category (i.e. pair-wise external indegree), the correlation between them is extremely low [Table 5]. At Tier-2 hierarchy, considering patent citations originating from any two external source subcategories towards the same destination subcategory (i.e. pair-wise external indegree), the correlation between them is low [Table 8] \\ \hline
5. &	Local processes (in subcategories) are driving local distributions even across hierarchical levels (for other categories)	& We study some random combinations across hierarchical levels of the USPCN, due to the high possibilities. We observe the prevalence of preferential attachment between groups across tiers [Table 6].  Further, patent citations originating from any two external Tier-2 source subcategories towards the same Tier-1 destination category is low [Table 7] \\ \hline
6. &	Percentage of vertices that are externally popular is low	& Percentage of vertices that have inter-community/ external in-links is low (Tier-1 range: 25\% to 28\% [Table 2]; Tier-2 range: 4\% to 29\% [Table 4]. We believe this to be another manifestation of locality. \\ \hline
7. &	Internal degree distribution type largely defines the global degree distribution type	& Refer [Table 2] and [Table 4]). We believe that this observation could be a result of the larger fraction of intra-community edges as compared to inter-community edges (refer Fig. 2 and Fig. 3) \\ \hline
8. &	Local distributions mostly follow a (truncated) power law.	& Truncated power law is the predominant best-fit distribution type for internal, external and global degree distributions, with some occurrences of power law and lognormal distribution types (refer [Table 2], [Table 4] and [Table 6]). \\ \hline
\end{tabular}
\end{table}

\section{Related Work}

A lot of past work has been done in the field of complex network modeling. The preferential attachment model introduced in [2] is one of the earliest attempts to explain the evolution of networks that obey power law degree distribution. Also known as the scale-free model, it can mimic degree distribution properties of diverse real-world social and biological networks, as well as the World Wide Web (WWW).

When considering models for community-based networks, the Stochastic Block Model (SBM) [3] is one of the most often used and cited generative model. Variants of the SBM model are described in [4] and [5]. {\it However, none of the SBM variants account for degree distributions at community level. Here, we show how the SBM matrix could be incorporated in building a network which satisfy certain local degree distribution constraints}.
The effect of “local events” in the evolution of networks has been studied in [8]. The term “local events” in [8] is defined as (i) new node additions, (ii) link formation due to new nodes and (iii) existing link rewirings. The authors describe how varying the frequency of occurrence of “local” (happening at a node) events leads to networks with (global) degree distribution following either a power law or an exponential distribution type. {\it In this paper, we are interested in networks with communities and local (in the group sense) behaviours}.

In extremely large networks, occurrence of local events (as defined in [8]) can happen in a state of information hiding (i.e. each vertex having a limited view of the overall network, to form its outgoing links). This aspect is discussed in further detail in [9]. In such a case, the authors discovered that the degree distribution follows a power law, but decays much faster, with an exponential truncation, referred to as a truncated power law distribution. While studying the effect of information hiding in the backdrop of local events, the focus in this paper remains at understanding the impact of such an environment to the global network degree distribution only. Further, in their analysis/model, a node can form an outgoing link to a randomly selected subset of nodes. {\it In the USPCN network with predefined categories, our analysis shows that each node forms most outgoing links within the group it belongs to, and only a few outside, rather than randomly}.

In [10], the authors extend the topic of local events and limited network state information to propose a local-world generative model, that allows for generation of a network which can transition between being a scale-free network and an exponential network. A more recent research based on extensions of the local-world generative model is described in [11], wherein the Neighbor-Preferential Growth (NPG) model is described. In both these studies, the “local-world” is defined to exist separately for each vertex, without any reference to communities. {\it On the contrary, our study shows that a vertex is influenced by the group it belongs to, when it comes to making links within and outside its group}.

\section{Conclusions}

In this paper, we analyzed local indegree distributions for the USPCN with its predefined categories and subcategories. We segregated the indegree of a vertex into internal indegree, from neighbors within (sub)category and external indegree, from neighbors in other (sub)categories. Our analysis at both levels suggest the predominant presence of local processes leading to global distributions. Each group has its independent perspective on how it views itself internally as well as externally. This may be due to the diversity of the subcategories within a category. We recognize that many of these findings are a consequence of the particular dataset we used. It would be very interesting to extend these studies to networks with organic communities (defined by modularity, for example) and networks with communities that allow a vertex to be a member of multiple communities. Such studies would shed light on the similarities and differences in community behaviour in various networks.

\end{document}